\documentstyle[preprint,aps]{revtex}
\begin {document}
\draft
\preprint{UCI-TR 96-7}
\title{Quark resonances and high E$_{\mbox{\rm t}}$ jets}
\author{Myron Bander\footnote{Electronic address: mbander@funth.ps.uci.edu}
}
\address{
Department of Physics, University of California, Irvine, California
92717}

\date{February\ \ \ 1996}
\maketitle
\begin{abstract}
Possible spin-3/2 quark resonances would have a
significant effect on high E$_{\mbox{\rm t}}$ jet production through
their contribution to the subprocess $q+{\bar q}\rightarrow g+g$. Such
enhancements are compared to a, recently reported, anomaly in inclusive
jet production at the CDF detector. 
\end{abstract}


The possible existence of spin 1/2 quark resonances has been looked
at, both theoretically \cite{resth} and experimentally \cite{resexp}.
In the latter two works excluded regions in the mass--coupling
constant plane were obtained; these analyses were based on the absence
of the direct production of such states. In this article we study the
effect of a possible {\em spin-3/2} quark resonance, $q^*$, on the
production rate of high E$_{\mbox{\rm t}}$ jets; the exchange of such a
particle in the reaction $q+{\bar q}\rightarrow g+g$ will enhance high
E$_{\mbox{\rm t}}$ gluon jet production. This
study is motivated by the recently observed excess of such
jets in 1.8 GeV $p{\bar p}$ collisions \cite{CDF}.

The $q-q^*-g$ interaction  Lagrangian with a minimum
number of derivatives is
\begin{equation}
{\cal L}=\frac{g_s\kappa}{2M^*}\left ({\bar q^*_\nu}\gamma_\lambda-
{\bar q^*_\lambda}\gamma_\nu\right )\frac{\lambda^\alpha}{2}q\,  
G_\alpha^{\nu\lambda}+\mbox{\rm h.c.}\, .
\end{equation}
$q^*_\nu$ is the Rarita-Schwinger field for a spin-3/2 particle,
$G_\alpha^{\nu\lambda}$ is the gluon field strength tensor, $M^*$ is
the resonance mass, $g_s$ is the strong coupling constant and $\kappa$
parameterizes the strength of this interaction. The quark $q$ and the
corresponding $q^*$ can represent either the $u$ or the $d$ quark. In
subsequent discussion we shall take the masses of $u^*, d^*$ to be
degenerate.

To order $\alpha_s$, the amplitude for $q+{\bar q}\rightarrow g+g$ is
given by the usual QCD diagrams, Fig.~1, and by the exchange of a
$q^*$, Fig.~2. For a spin-3/2 $q^*$ the amplitude due to latter will
grow by one power of ${\hat s}$, the quark-antiquark center of mass
energy squared, faster than the amplitude described by Fig.~1. Such
growth cannot go on indefinitely as at high enough ${\hat s}$ an
exchange of a spin-3/2 particle will violate unitarity.  However,
depending on the strengths of various couplings, this increase can
persist to large values of ${\hat s}$ values, typically up to ${\hat
s}\sim M^{*2}/(\alpha_s\kappa^2)$; this limit is well beyond the range
of ${\hat s}$ needed at present.

The effect of such exchanges on E$_{\mbox{\rm t}}$ distributions can
be seen in Fig.~3, where we show the transverse energy distribution,
at rapidity $y=0$, for gluon jets from the parton reactions $u+{\bar
u}\rightarrow g+g$ plus $d+{\bar d}\rightarrow g+g$. In all
calculations presented here we used the MRSA'~\cite{MRS} parton
distribution functions ($\Lambda_{QCD}=231$ MeV) with renormalization
scale $\mu$=E$_{\mbox{\rm t}}/2$. Values for $M^*$, the mass of the
spin-3/2 resonance and for its coupling strength, $\kappa$ (cf. Eq.~
(1)), are the ones used below in comparing to the data of
Ref.~\cite{CDF}. For E$_{\mbox{\rm t}}< 200$ GeV the contribution the
exchange of the spin-3/2 resonances is small; this exchange rapidly
dominates the QCD contributions for larger values of E$_{\mbox{\rm
t}}$.
 
In Fig.~4 we plot the sum of the square of the amplitude due to the
exchange of a spin-3/2 particle (Fig.~2) and the interference of this
and the QCD amplitude (Fig.~1) as a percentage of the corresponding
next to leading order QCD calculation \cite{NLO} of the single jet inclusive
cross section and compare to the experimental results of Ref.
\cite{CDF}; the extra contributions is averaged over the
pseudorapidity range $0.1\le\eta\le 0.7$. We restricted the comparison
to $\kappa\le 1.0$. For $\kappa=1.0$, $M^*=400$ GeV is largest
resonance mass that the data can accommodate; for smaller $M^*$ we
restricted the comparison to $M^*\ge 100$ GeV and for the latter we
obtain a good fit with $\kappa=0.13$.

As mentioned earlier, restrictions on masses and couplings of spin-1/2
quark resonances exist \cite{resth,resexp}. As we are dealing with a
spin-3/2 system these restrictions cannot be taken over directly;
namely, restrictions on $f_s$ of Refs. \cite{resth,resexp} do not
translate into restrictions on $\kappa$. These analyses also assume
that analogous couplings of spin-1/2 $q^*$'s to photons and $W$'s are
proportional to $f_s$. The only unambiguous carry over of the spin-1/2
analysis to the spin-3/2 case is for the process of quark-gluon fusion
where we find \cite{KuZe} $\sigma(g+q\rightarrow q^*_{s=3/2})=
0.5\, \sigma(g+q\rightarrow q^*_{s=1/2})$ and thus it is somewhat harder
to produce the spin-3/2 state.  Should we anyway make such a
comparison, it is comforting that these are not far off; $M^*=100$
GeV, $f_s=0.13$ is in the allowed range, while for $f_s=1.0$ the
spin-1/2 analysis would limit $M^*$ to $M^*>560$ GeV, again not far
away from the value $M^*=400$ GeV used in this work. A reanalysis of
the data of Ref. \cite{resexp} using spin-3/2 excited quarks is
needed.

In this study we have shown that spin-3/2 quark resonances can account
for the observed large high E$_{\mbox{\rm t}}$ cross section. 

I would like to thank Dr. Walter Giele for discussions and for
providing me with parton structure functions.

\nobreak 

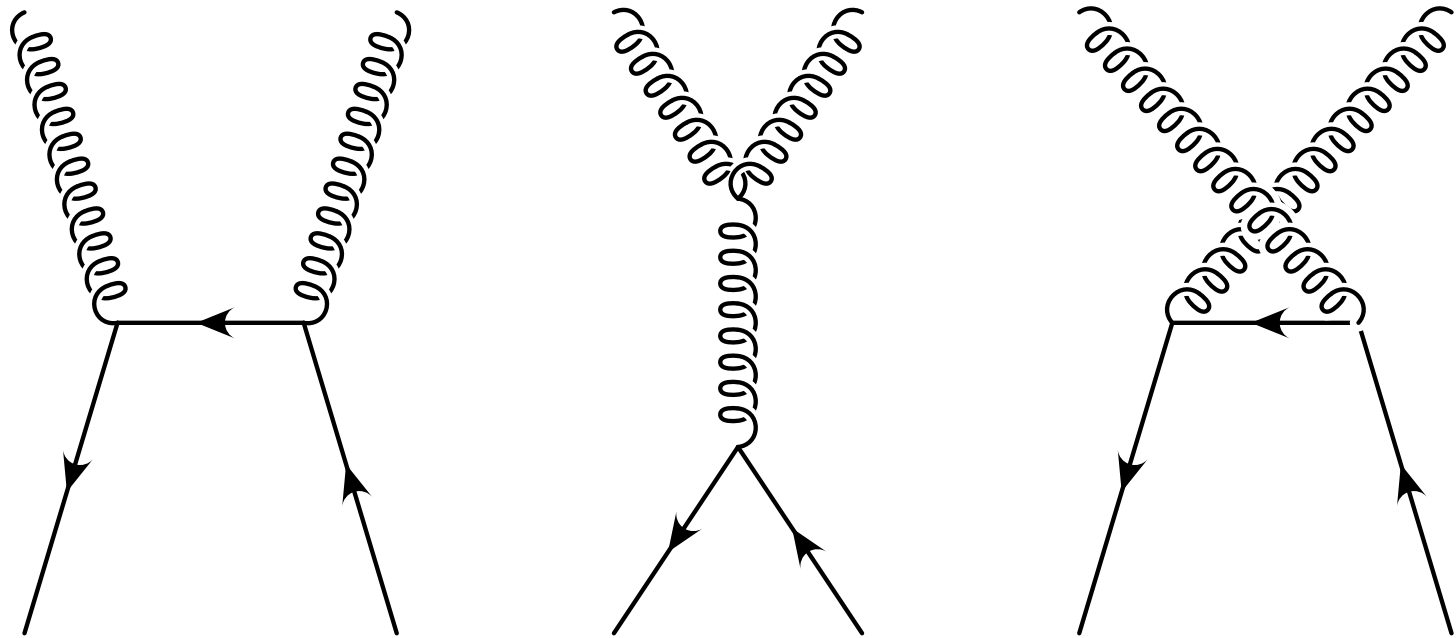
\begin{figure}
\caption{ QCD Feynman diagrams for the process $q+{\bar q}\rightarrow
g+g$. }\label{fig. 1}
\end{figure}

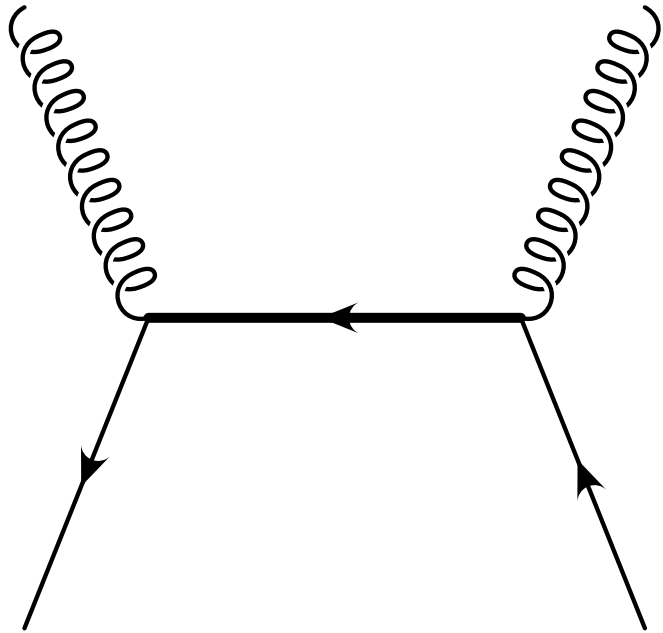
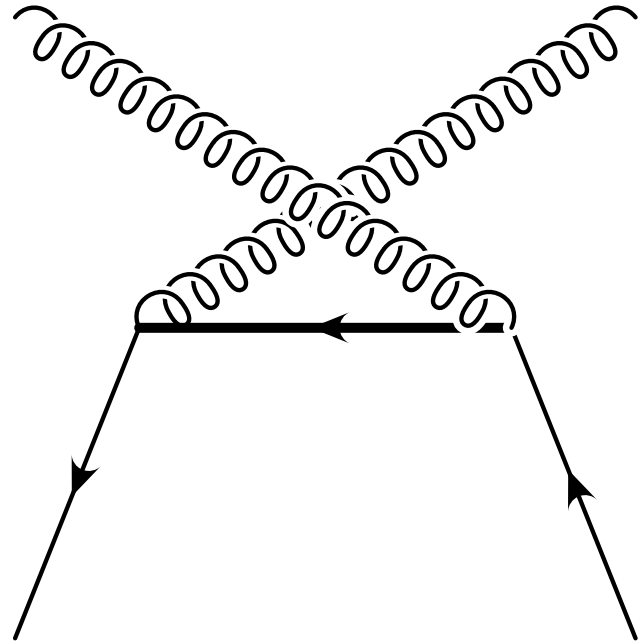
\begin{figure}
\caption{ Feynman diagrams for the exchange of a spin-3/2 quark
resonance in the process $q+{\bar q}\rightarrow g+g$. }\label{fig. 2}
\end{figure}

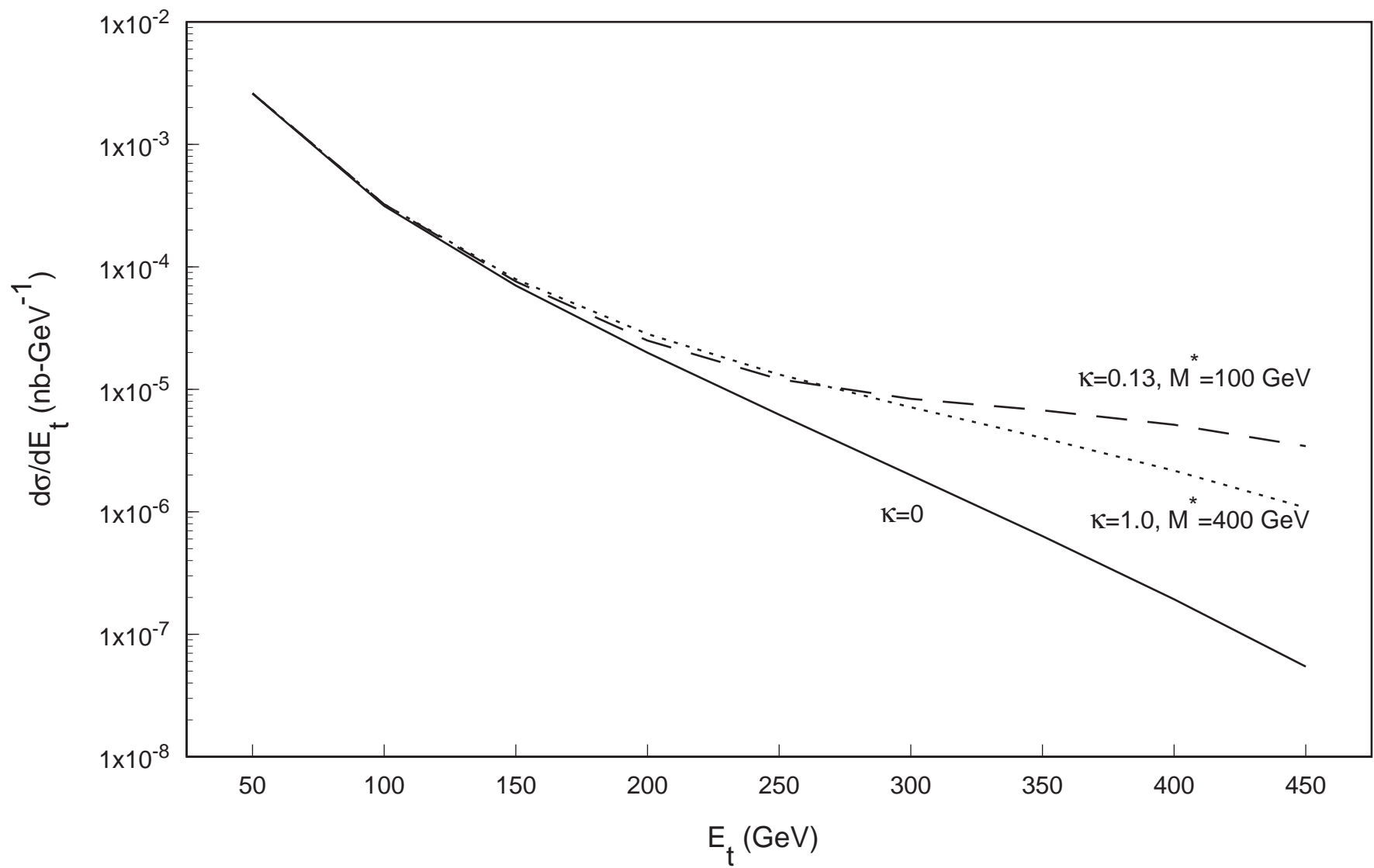
\begin{figure}
\caption{ Transverse energy distribution for $p+{\bar p}\rightarrow
g+g$, at 1.8 TeV center of mass energy and rapidity $y=0$, from the
subprocess $q+{\bar q}\rightarrow g+g$, for several values of the
spin-3/2 quark resonance mass $M^*$ and coupling strength
$\kappa$ (cf. Eq.~1); the solid line is for $\kappa=0$, the dashed one
for $\kappa=0.13,\ M^*=100$ GeV and the dotted one for $\kappa=1.0,\
M^*=400$ GeV. }\label{fig. 3}  
\end{figure}

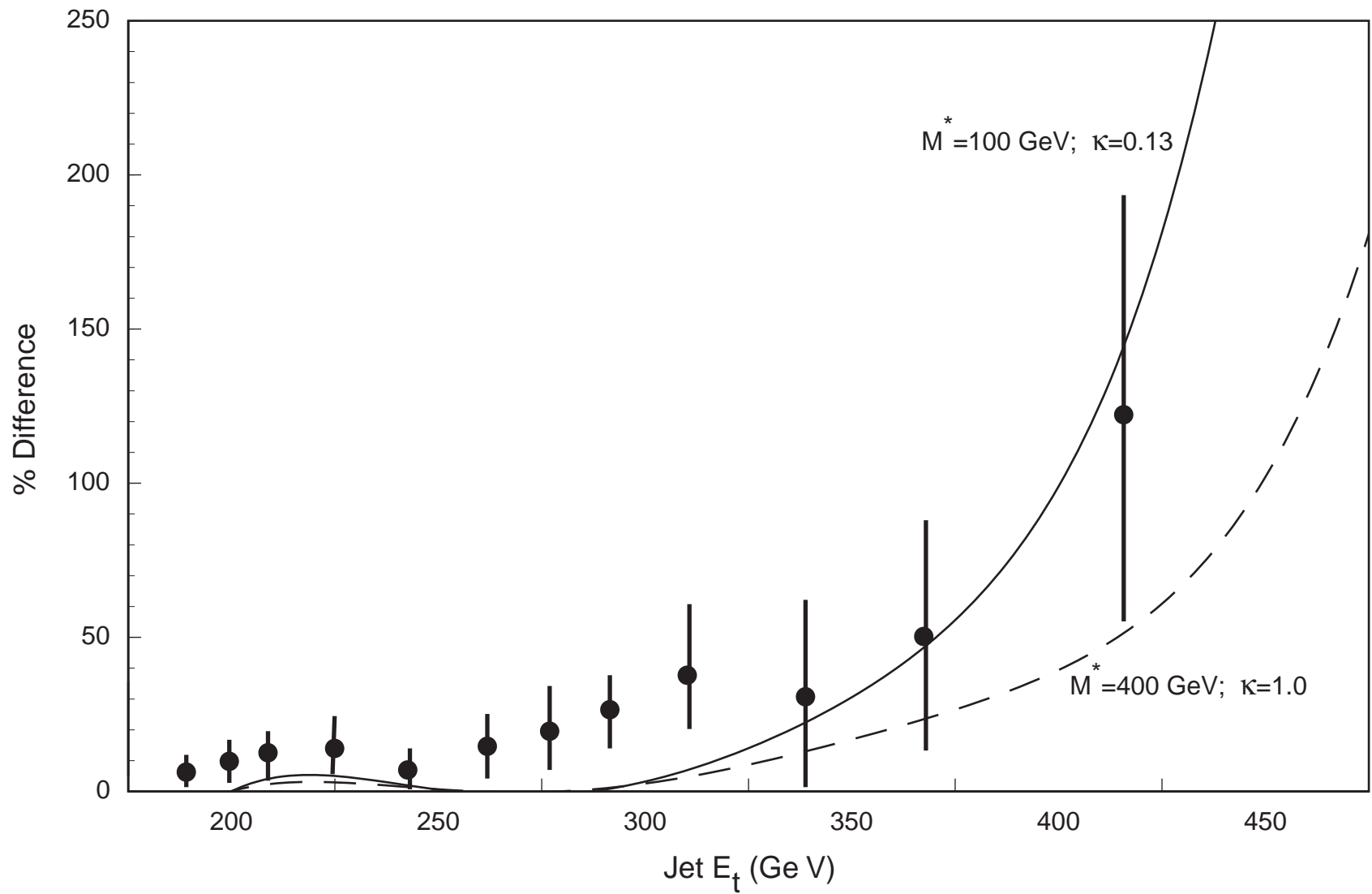
\begin{figure}
\caption{The percent difference between the experimental inclusive jet 
transverse energy distribution (Ref. 3), the contributions 
due to the exchange of a spin-3/2 quark resonance and pure QCD
predictions used in Ref. 3; solid curve is for $\kappa=0.13,\
M^*=100$ GeV and the dashed one for $\kappa=1.0,\ M^*=400$
GeV. }
\label{fig. 4}
\end{figure}
\end{document}